# Extending Jacobian matrix in proving stability for nonlinear systems with one equilibrium point such as compressor

**Seyed Mohammad Hosseindokht**

*Department of Electrical Engineering*
*Iran University of Science and Technology*
*Tehran, Iran*

**SamanehAlsadat Saeedinia**

*Department of Electrical Engineering*
*Iran University of Science and Technology*
*Tehran, Iran*

## Abstract

Global stability of the systems has always been vital of importance; however, this concept has not yet been sufficiently developed for the nonlinear systems. This paper extends the Jacobian matrix so that this method be able to seek the criteria to ensure global stability for a special class of nonlinear systems. In this regard, we propose a new analysis method that utilizes the Jacobian matrix concept, integrating with the characteristics of the negative eigenvalues to analyze the global stability of the nonlinear systems with only one equilibrium point. Also, the positive eigenvalue to analyze the global instability of the nonlinear systems with only one equilibrium point. Some theorems such as Hartman-Grobman and Popov criteria can prove this claim. To this end, several examples and a benchmark systems have been intended to evaluate the efficiency of the proposed method. Results indicate the high potential of the proposed approach in order to develop the global stability analysis. The nonlinear compressor model, categorized in this extensive class, is also investigated as a well-known industrial system besides other several examples. The outcomes demonstrate that extended Jacobian stability analysis can ensure global stability for this class of nonlinear systems under some spatial conditions, discussed in this paper.

# 1. Introduction

Having studied the last two- decades of literature on the nonlinear systems, we perceive that their global stability has highly grasped the researchers' attention [1, 2, 3]. In this regard, a lot of significant systematic methods have been introduced to extend the attraction regions [1, 3]. To this end, various methods have been developed for different nonlinear system classes, and correspondingly their stability has been analysed [1, 3]. These methods can be basically classified into three main categories; namely, equilibria-dependent stability analysis, orbital stability of the output trajectory, and structural stability. The stability analysis, based on equilibrium state, has been embarked as early as 1644 [4].

The stability theory of Lyapunov is one of the highlighted stability analysis methods, extended by the great numbers of researchers for the various classes of the nonlinear systems [5, 6, 7]. One of the eminent results of this theory is the Jacobian conjecture theory which has been established and investigated in only two dimensional systems with the aim of finding auxiliary boundary value problem to ensure the global asymptotic stability [8]. This theory states that x=0 is an asymptotic stable equilibrium point for a two-dimensional system of ordinary differential equations if the trace of the Jacobian matrix is negative and its determinant is positive, which implies having negative real-part eigenvalues and system stability.

This idea developed to n dimensional first order differential systems, expressing that if the real part of the eigenvalues of the J(x) in critical point O are negative, then O is asymptotically stable. In this regard, Orbital stability states [9] that if there is a close set of p-periodic solutions for an ordinary autonomous system in which any perturbation near the equilibrium in the p-periodic solution orbit, stays on that neighbourhood; therefore, p-periodic solution trajectory, $\varphi_t(x_0)$), is said to be orbitally stable.

Oppose to the both described stability theorem, it would be useful to note the linear instability theory which can be utilized to disprove stability. This theory expresses that for an autonomous system, if Jacobian Matrix *(J(x=x\*=0))* has at least one positive eigenvalue, then *x\* =0* is unstable [4], which implies non-positive eigenvalue is the essential, not sufficient, condition for stability analysis of a system.

Jacobian matrix method analyses stability of the systems based on linearization and determining eigenvalues in the equilibrium points. In this method, stability is investigated at equilibrium points locally not globally. Thus, stability of other points, located out of equilibrium points, is unknown. It means that in nonlinear systems with several equilibrium points, stability of the state trajectories is indefinite. However, for the nonlinear system with only one equilibrium point, Jacobian matrix method can be utilized to prove global stability under some circumstances. On the other hand, if the system has one and only one unstable focus, it is unstable at any initial operation point, and the states of the system diverge from equilibrium point to infinite. For the system with more than one equilibrium points, Jacobian matrix is not enough to analyse the stability of the system; hence, phase plane should be applied to describe stability of the system, globally. Inspired by both the orbital and Jacobian Conjecture stability concepts, this paper proposes a new stability analysis approach for a spatial class of nonlinear systems, with one feasible equilibrium point [10]. To this class, this paper proposes a new method, in which the Jacobin matrix stability approach has been extended so that the global stability criteria of the relevant system are determined. In another word, regardless of the state-space model dimension, if there is a single feasible stable equilibrium point, we can conclude that if the system is stable at its unique equilibrium point, the system can be globally stable under spatial conditions. Otherwise, the system is unstable or locally stable. This subject is not still a theorem, but by the time, there are no counter examples. The proposed stability analysis method is on the basis of calculating the eigenvalue of the system at the unique equilibrium point and solving the differential equations to provide the sufficient condition for global stability of the system.

Lyuponov theorem for analysis of nonlinear system with one equilibrium point is very useful, but this theorem is sufficient condition for stability. In the other hand, Lyapunov theorem and other methods such as LPV (linear parameter varying), nonlinear embedding and LMI(linear matrix inequality) can prove stability only and instability at all. Our claim (extending Jacobian matrix method) can prove stability and instability clearly. In the other hand, extending Jacobian matrix method is necessary and sufficient condition for stability and this method can prove instability. Moreover, Jacobian matrix is very fast and easy fundamentally and mathematically.

Hartman–Grobman theorem and its related book [11] can qualify and confirm our claim for global stability for simple nonlinear system.

Moreover, Popov theorem [12] is very powerful theorem for proving our claim in this paper. In this paper, mentioned claim is proved with Popov theorem for global asymptotic stability and instability. This necessary and sufficient condition.

In order to expand the issue, the stability of several numerical examples has been evaluated. To indicate the engineering aspect of the suggested solution for the stability analysis of the mentioned class, the well-known Greitzer model which is used for estimation of the coupled behaviour of compressors' pressure and flow dynamics, has been analysed. The result indicates the suggested solution brings about the less conservative global stability condition for the system in comparison to the relevant studies [12, 13].

## 2. Extended Jacobian Conjecture for n dimensional ordinary system

This paper proposes a methodology to analyse the global stability for n-dimensional ordinary system by solving differential equation approaches for several nonlinear classes, possessing only one equilibrium point. To this end, consider an autonomous system as follow:

$$\dot{x} = f(x) \quad , \quad x(t_0) = x_0 \in R^n \tag{1}$$

The Jacobian matrix is calculated as below:

$$J(x) = \frac{\partial f(x)}{\partial x} = \left[\frac{\partial f(x)}{\partial x_1} \cdots \frac{\partial f(x)}{\partial x_n}\right] \tag{2}$$

Eigenvalues of the J(x) are calculated at the equilibrium point.

$$\det(\lambda I_{n*n} - J) = 0 \quad , roots: \lambda_1, \lambda_2, \dots, \lambda_n \tag{3}$$

Where *det(.)* denotes the determinant of a matrix. Generally, Jacobian Matrix stability analysis is able to indicate the local stability in the nonlinear systems, however, due to linearization of the system, global stability of the system cannot be guaranteed even if the system possesses one and only one stable equilibrium point. Since the parametric calculation of all eigenvalues is computationally expensive, this paper postulates that calculation of the eigenvalues of the mentioned class of nonlinear systems, only at their unique equilibrium point is sufficient for their stability analysis. So, if all the real parts of the eigenvalues are totally negative, global stability can be ensured. In contrast, the existence of even one positive eigenvalue indicates the instability of the system. Otherwise, the system is conditionally stable. In fact, due to the existence of only one equilibrium point in the nonlinear system, we can claim that Jacobian Matrix Analysis at the equilibrium point is sufficient for stability analysis. This matter distinguishes the proposed solution from other Jacobian Matrix Stability analysing methods. To extend and clarify the proposed methodology, several examples of nonlinear systems are described in the following sections. In continue, solving differential equations and figures of states in several initial values can help to prove this claim.

## 3. Hartman–Grobman theorem

In this part Hartman–Grobman theorem is displayed, because this theorem is related to topologically conjugate or mathematic and geometric mapping between nonlinear system and linearization.

In mathematics, in the study of dynamical systems, the Hartman–Grobman theorem or linearization theorem is a theorem about the local behaviour of dynamical systems in the neighbourhood of a hyperbolic equilibrium point. It asserts that linearization—a natural simplification of the system—is effective in predicting qualitative patterns of behaviour. The theorem owes its name to Philip Hartman and David M. Grobman.

The theorem states that the behaviour of a dynamical system in a domain near a hyperbolic equilibrium point is qualitatively the same as the behaviour of its linearization near this equilibrium point, where hyperbolicity means that no eigenvalue of the linearization has real part equal to zero. Therefore, when dealing with such dynamical systems one can use the simpler linearization of the system to analyse its behaviour around equilibrium [11].

When nonlinear system has only one equilibrium point, analysing stability with linearization around equilibrium point is global. In book [11], global stability of nonlinear system with one equilibrium point is mentioned. In the other hand, nonlinear system with only one equilibrium point is simple nonlinear system and analysing stability with linearization or Jacobian matrix is global stability.

## 4. Popov theorem (proving stability)

In this part proving stability with Popov theorem will be described. Based on this theorem, proving stability is global for any nonlinear system [14].

Popov theorem describes that In nonlinear control and stability theory, the Popov criterion, introduced by Vasile M. Popov, is a stability criterion used to determine the absolute stability of certain nonlinear systems, provided their nonlinearity meets an open-sector condition. Unlike the circle criterion, which can be used for nonlinear time-varying systems, the Popov criterion is specifically limited to autonomous (time-invariant) systems.

Sub-class of Popov theorem are in this way:

$$\dot{x} = Ax + bu$$
$$y = cx + du$$
$$u = -\varphi(y) \quad (4)$$

Here $x \in R^n$ and u and y are scalar variables, while A, b, c and d have compatible dimensions. The nonlinear function $\varphi: R \to R$ is a time-invariant nonlinearity that belongs to the open sector $(0,\infty)$ meaning $\varphi(0) = 0$ and $y\varphi(y) > 0$ (5) for all y≠0 such that $0 < \varphi(y) < k$.

It is important to note that the system analyzed by Popov has a pole at the origin and lacks direct input-output transmission, with the transfer function from u to y given by

$$H(s) = \frac{d}{s} + c(sI - A)^{-1}b \quad (6)$$

There are another criteria or conditions for fulfilment of this theorem:

A: Hurwitz(all eigenvalues should be negative), (A,b): controllable,

(A,c): observable, $\varphi \in (0, \infty)$.

The system is globally asymptotically stable if there exists a positive number γ>0 such that:

$$\inf_{w \in R} Re\{(1 + jw\gamma)H(jw)\} > 0 \quad \text{or} \quad \frac{1}{k} + Re(H(jw)) - w\gamma Im(H(jw)) > 0 \quad (7)$$

Mentioned statements were about description of Popov theorem, but it is essential description of relation between Popov theorem and our claim in this paper. In the other hand, criteria and conditions should be satisfied.

In autonomous nonlinear model there is $\dot{x} = f(x)$ and with Taylor expansion around equilibrium point we have:

$$\dot{x} = f(x_0) + J(x_0).x + H.O.T (higher\ order\ terms) \quad (8)$$

Based on our assumption eigenvalues of Jacobian matrix is negative and Taylor expansion is around equilibrium point then $\dot{x}(x_0)=0$, $f(x_0)=0$ then

$$\dot{x} = J(x_0).x + H.O.T (higher\ order\ terms) \quad (9)$$

We can consider similarity between our claim and Popov condition.

$$\dot{x} = J(x_0).x + H.O.T (higher\ order\ terms) \quad , y = x \quad (10)$$

Based on relation between (4) and (10), we have:

A=$J(x_0)$ , b=1 , c=1 , d=0 , u= H.O.T= $-\varphi(x)$ (11)

We should prove if A=$J(x_0)$ is Hurwitz or all eigenvalues of Jacobian around equilibrium point is negative then system is stable. Moreover, We should prove if A=$J(x_0)$ is not Hurwitz or all eigenvalues of Jacobian around equilibrium point is not negative then system is unstable.

Based on (6) and (11) we have: $\quad H(jw) = \frac{1}{jw-A}$ (12)

Based on (7) we have: $\quad \frac{1}{k} + \frac{-A}{w^2+A^2} + \frac{\gamma w^2}{w^2+A^2}$ (13)

we can consider k infinite, because in every way this limitation and boundary is true or we can choose every number for k. We can choose k infinite for simplifying. And then (13) will be $\gamma w^2 > eig(A)$. This relation is true in any way and any dimension of A, because all eigenvalues of A are negative and there is $\gamma > 0$ for satisfying (13). With choosing number for k, we have:
$w^2(1 + k\gamma) + A^2 > kA$

This inequality is true based on every $\gamma, k > 0$.

Another subject should be described for completing Popov theorem and proving asymptotic stability and instability.

Because of being only one equilibrium point in our assumption then in (10), $\varphi(x)$ is zero only in one equilibrium point and because there is not another equilibrium point. Then around only one equilibrium point, $\dot{x}$ should be negative for $x > 0$ and positive for $0 > x$ until nonlinear system around equilibrium point is stable, because all eigenvalues of Jacobian is negative.

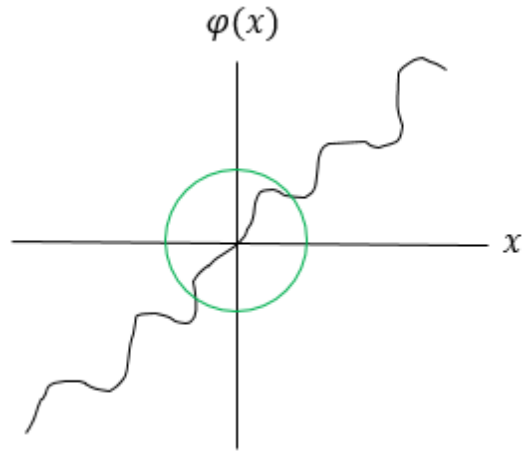

*Figure 1 changing φ(x) based on x*

Because around equilibrium point $\dot{x}$ should be negative then $\varphi(x) > 0$ and because there is only one equilibrium point, then $\varphi(x)$ is zero only in one equilibrium point and in another points $\varphi(x)$ is not zero then $\varphi(x)$ don't have changing sign and in result $\varphi(x) > 0$ for x > 0 . Finally, all conditions and assumptions of Popov theorem is fulfilled and nonlinear system is global asymptotic stable. Moreover, Popov theorem is sufficient and necessary condition and if all eigenvalues of Jacobian matrix is not negative then system is global unstable.

## 5. Applications of the proposed method in global stability analysis

In this section, some applications of the proposed stability analysis method are epitomized.

### 5.1. Example 1:

Consider the following example:

$$\dot{x} = \sqrt{x} - 1 \tag{14}$$

The equilibrium point of the system is x=1. Jacobian Matrix at the equilibrium point is $J = \left.\dfrac{1}{2\sqrt{x}}\right|_{x=1} = \dfrac{1}{2}$. Therefore, the eigenvalue of the system is positive, and consequently, system is unstable. Solution of the equation can also proof this matter; however, the Linear instability theorem is sufficient to conclude instability of the system. Solution of the system is given below:

$$u = \sqrt{x}$$

$$\dot{x} = \sqrt{x} - 1 \to 2u\dot{u} = u - 1 \to \dfrac{2u}{u-1} du = dt$$

$$\to \left(2 + \dfrac{2}{u-1}\right) du = dt \xrightarrow{\int} u + \ln(u-1) = \dfrac{t+k_0}{2} \to \ln(u-1) = -u + 0.5t + k_2 \quad (15)$$

$$\to u = 1 + \eta e^{0.5(t-2u)} \to x = \left(1 + \eta e^{0.5(t-2\sqrt{x})}\right)^2$$

$$\lim_{\substack{t \to \infty \\ x \geq 0}} \left(1 + \eta e^{+0.5(t-2\sqrt{x})}\right)^2 = \infty$$

So, the system is unstable.

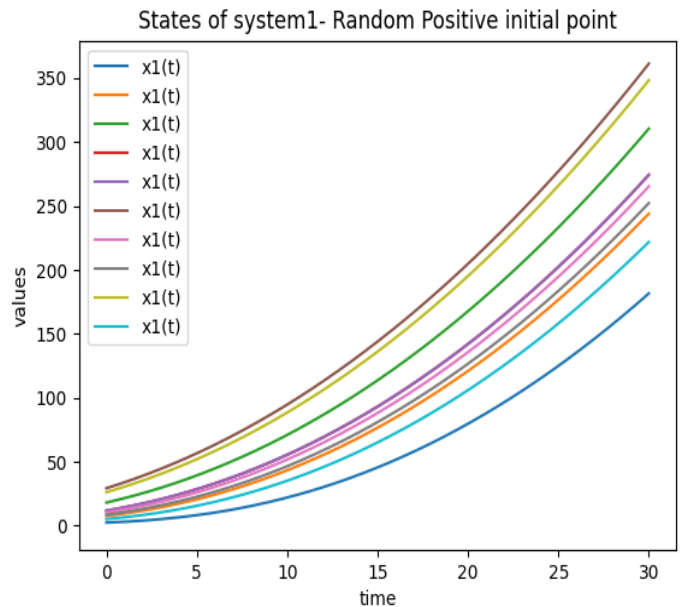

Figure 2 States of the system 1 at random initial states

Figure 2 reveals that all the states diverges in random initial conditions.

## 5.2. Example 2:

Consider the second example of nonlinear system as follows:

$$system2: \dot{x} = -\sqrt{x} + 1, \quad x > 0, \quad x \in \Re \quad (16)$$

Equilibrium point of this system is x=1. Jacobian Matrix of the system is calculated below:

$$J = -\frac{1}{2\sqrt{x}}\bigg|_{x=1} = -\frac{1}{2} \quad (17)$$

According to the calculated negative eigenvalue of the system $\lambda = -\frac{1}{2}$, we can conclude that the system is stable, and due to the existence of only one equilibrium point and holding second condition (Eq.16), the paper claims that the system is globally stable.

$$J^{-1}\dot{J} \square \; \varepsilon(X-X_e)^T\left((X-X_e)\left((X-X_e)^T\right)\right)^{-1} \;,\; 0 < \varepsilon \square \; 1 \rightarrow \frac{-1}{8x} < \varepsilon \frac{1}{x} \quad True \quad for \quad x > 0, \varepsilon > 0 \quad (18)$$

To investigate the accuracy of this postulate, the state trajectory and phase portrait of the system at random positive initial points *are* given in Fig.3, respectively.

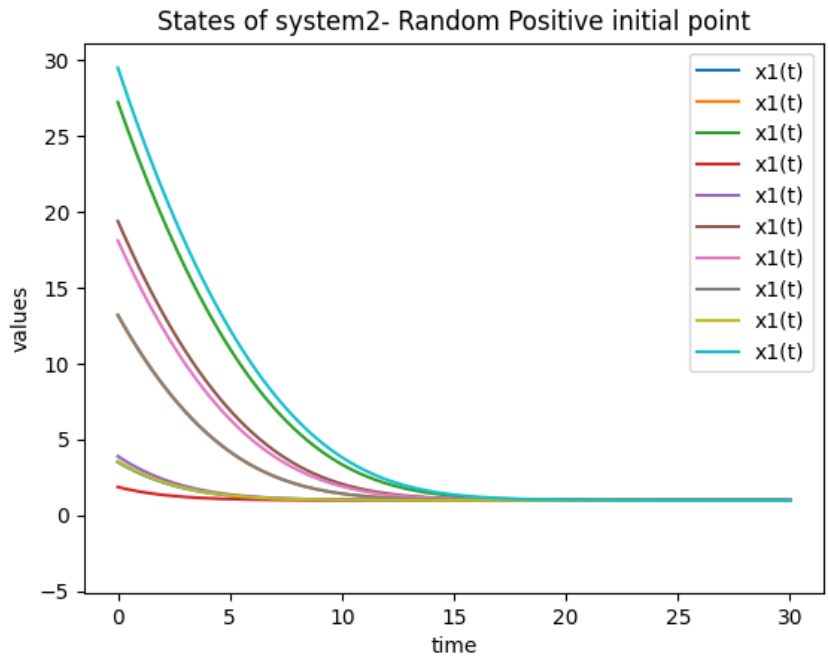

*Figure 3 state trajectory of the system2 at 10 random positive initial points*

Cleary, all the given trajectories of the system converge to 1, at the equilibrium point. For more analysis, the phase portrait of the system is given below:

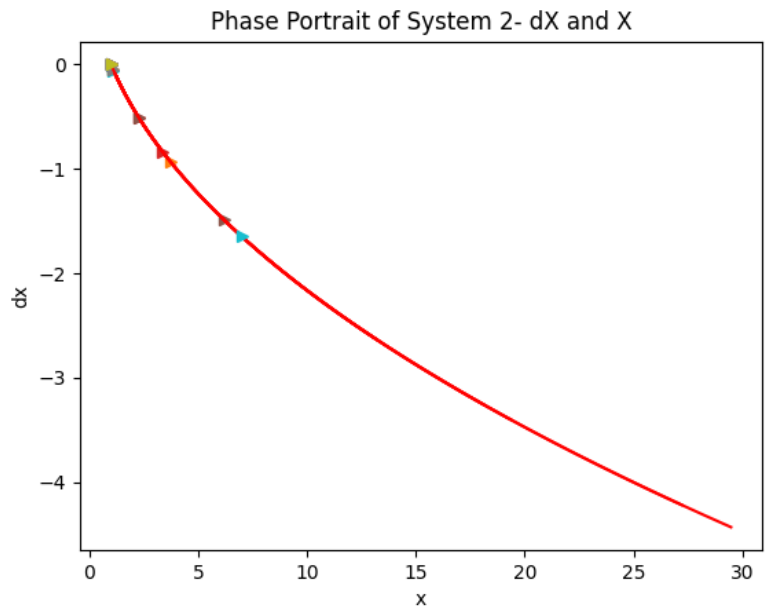

*Figure 4 Phase Portrait of the system 2*

As shown in Fig.4, the fractional system example is globally stable, due to the convergence of all trajectories to zeros. In addition, the solution of the exampled ordinary differential equation (ode), calculated below, indicates convergence of the system in all initial states, and therefore, it implies global stability.

$$u = \sqrt{x}$$

$$\dot{x} + \sqrt{x} - 1 = 0 \rightarrow 2u\dot{u} + u - 1 = 0 \rightarrow \frac{2u}{1-u} du = dt$$

$$\rightarrow \left(-2 + \frac{2}{1-u}\right) du = dt \rightarrow -2u - 2\ln(1-u) + k_0 = t + k_1 \rightarrow 1 - u = e^{-0.5(t+2u+c_0)} \quad (19)$$

$$\rightarrow u = 1 - \eta e^{-0.5(t+2u)} \rightarrow x = \left(1 - \eta e^{-0.5(t+2\sqrt{x})}\right)^2$$

$$\lim_{\substack{t \to \infty \\ x \geq 0}} \left(1 - \eta e^{-0.5(t+2\sqrt{x})}\right)^2 = 1$$

The infinite limit of the calculated states implies convergence to the unique equilibrium point of the system. Although it is assumed a real value for x, and system (x>0), considering complex values does not change the stability condition as indicted follows:

$$\lim_{\substack{t \to \infty \\ x < 0}} \left(1 - \eta e^{-0.5(t)+i2\sqrt{|x|}}\right)^2 = \lim_{t \to \infty}\left(1 - \eta e^{-0.5(t)} e^{i(2\sqrt{|x|})}_{\theta}\right)^2 = 1, \quad \left|e^{i(2\sqrt{|x|})}_{\theta}\right| = 1 \quad (20)$$

## 5.3. Example 3:

Consider the following state space equations of the system:

$$system3: \begin{cases} \dot{x}_1 = \dfrac{1-x_1^3}{3} \\ \dot{x}_2 = -(x_1^2 + 1)(x_2 - 1) \end{cases} \quad (21)$$

Equilibrium point of the given system is (1,1). Thus, Jacobian Matrix of the system at the equilibrium point is calculated as follows:

$$J = \begin{bmatrix} -x_1^2 & 0 \\ 2x_1(x_2-1) & -(x_1^2+1) \end{bmatrix}\bigg|_{(1,1)} = \begin{bmatrix} -1 & 0 \\ 0 & -2 \end{bmatrix} \quad (22)$$

Eigenvalues of the system are respectively (-1,-2). Consequently, negative eigenvalues imply on stability of the system. According to our conjecture, the sign of nonzero eigenvalues at the unique equilibrium point can be sufficient to ensure the global stability of the system, which this issue can be ensured by Markus-yamabe Conjecture, proved by [14] because of possessing two dimensions and having nonzero negative eigenvalues $-x_1^2$ and $-(x_1^2+1)$ for all points. Therefore, system is inherently globally stable. Since Markus-Yamabe conjecture can only develop for two-dimensional system with strictly negative eigenvalues for all x, our proposed conjecture can ensure the global stability, regardless of dimensions [15, 16].

To evaluate this issue, consider following system higher dimensional examples.

## 5.4. Example 4:

Consider the following nonlinear system:

$$\dot{X} = f(x_1, x_2, x_3)$$
$$X = \begin{bmatrix} x_1 & x_2 & x_3 \end{bmatrix}^T \tag{23}$$
$$system1: \begin{cases} \dot{x}_1 = x_3 \\ \dot{x}_2 = (x_2 - x_3)^2 \\ \dot{x}_3 = x_1 - 1 + x_2 \end{cases}$$

With only one equilibrium point at (1,0,0).

As the first step, Jacobian Matrix of the system is described as follows:

$$J = \frac{\partial f}{\partial X} = \begin{bmatrix} \frac{\partial f}{\partial x_1} & \frac{\partial f}{\partial x_2} & \frac{\partial f}{\partial x_3} \end{bmatrix} = \begin{bmatrix} 0 & 0 & 1 \\ 0 & 2(x_2 - x_3) & -2(x_2 - x_3) \\ 1 & 1 & 0 \end{bmatrix}_{(1,0,0)} = \begin{bmatrix} 0 & 0 & 1 \\ 0 & 0 & 0 \\ 1 & 1 & 0 \end{bmatrix} \tag{24}$$

The eigenvalues of the Jacobian matrix are calculated by solving the following equation:

$$\det(sI - J) = 0 \tag{25}$$

Where I is the identity matrix, $s$ is a Laplace variable, indicating Eigen values or poles of the system, and J is the Jacobian matrix. By solving Eq.25, Eigen values are respectively 0, 1 and -1 at the equilibrium point. This matter implies on instability of the system.

Figure 5 demonstrates the parametric eigenvalues' surfaces, indicating stability criteria of the system.

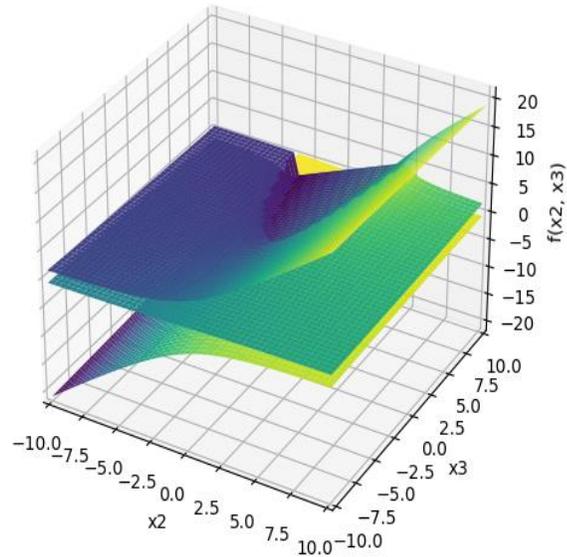

*Figure 5 Parametric eigenvalues of the System 1*

Figure 5 indicates that one of the eigenvalue surfaces is positive and consequently the system is unstable. Since Jacobian Matrix is generally utilized for stability analysis of the system, the surface indicates that for all initial states and working points, the stone of the eigenvalues of the system is positive, and therefore, the system is totally unstable. For more clarification, the states of the system are indicated in Figure 5 at the initial point [1.5,0,0].

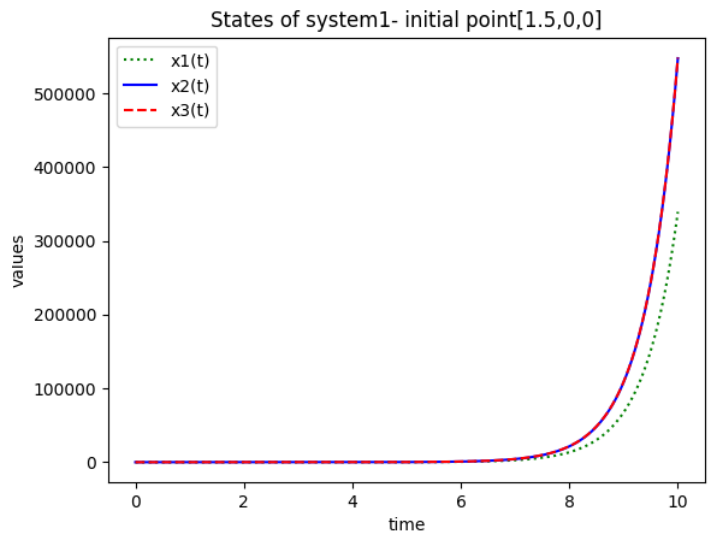

*Figure 6 State trajectories of system 1 at initial point[1.5,0,0]- Unstable example*

Figure 6 indicates that all the states of the system are unstable. As it is shown in Fig.6, a positive eigenvalue surface is relevant to x2, which indicates high-speed divergence. For further analysis, the phase portraits of the system are also demonstrated in Fig 7(a, b, and c).

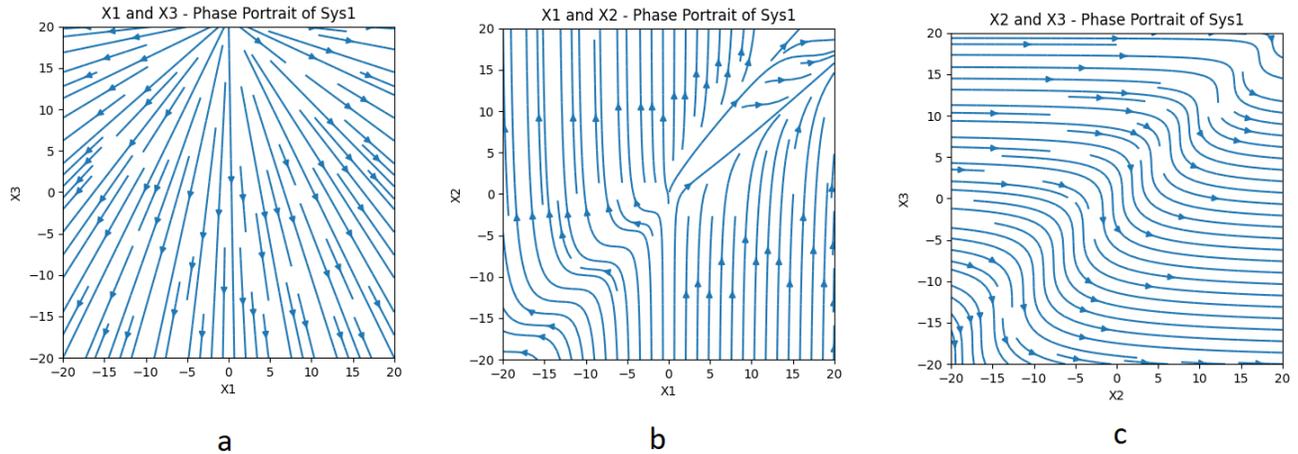

Figure 7 Phase Portrait of System 1- a) x1-x3  b)x1-x2   c)x2-x3

Figure 7 reveals that all the states' trajectories diverge, and consequently system is unstable. Figure 7.a reveals that there is a linear relationship between states $x_3$ and $x_1$ in all trajectories. Figure 7.b indicates a bifurcation nearby the equilibrium point, and Fig.7c shows a saddle point relationship on the line $x_2=x_3$.

## 6. Stability analysis of Nonlinear Compressor system and surge

### 6.1. Stable and unstable regions

The well-known Greitzer model is used for the estimation of the coupled behaviour of compressors' pressure and flow dynamics. Due to the vastly utilizing of the compressors in industries such as refineries, chemical and petroleum plants, and large-scale refrigeration, their stability, and performance, have always been a concern.

Surge, as one of the events which trigger instability, manifests when a compressor operates in a small mass flow rate at the constant speed or a given pressure ratio [17]. High pressure in the outlet and low flow inertia result in an intense reverse flow. Despite reverse flow, the rotor transfers the energy and forces the compressor to return to normal forward flow and higher pressure in the outlet. However, in the same throttling condition, outlet pressure and reverse flow once again overcome inlet flow. If the problem, leading to the original surge event, is not addressed, the surge cycles will repeat [18]. This cyclic process continues very quickly, and simultaneously amplitude of several thermodynamic or aerodynamic parameters drastically oscillates [19]. The instabilities and parameter oscillations are shown in Fig 8. These oscillations terribly can damage the compressor and even can induce vibration to another component.

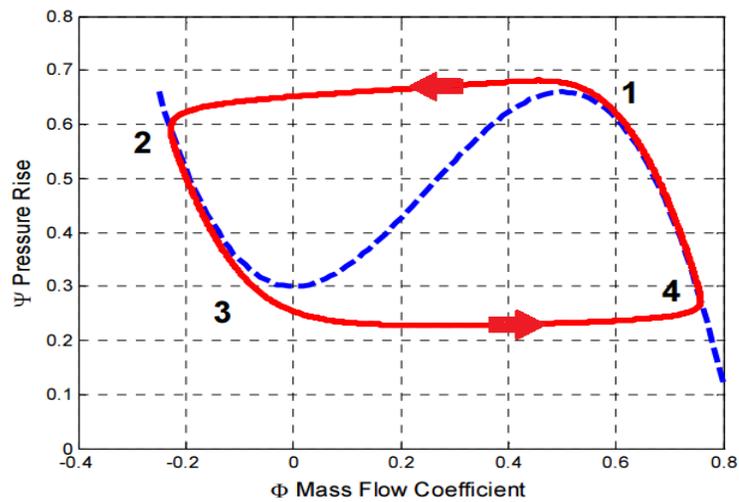

*Figure 8 Characteristic curves and surge phenomenon cycle*

The issue can even threaten and disturb the safety and durability of the internal component of the system [20]. Therefore, the prediction of instabilities, and consequently ability to determine surge point and stable operation of a compressor performance map are essential. However, there are a great number of researches in the field of analysing the stability of the compressors or designing the compressors' anti-surge controller controller [20, 21, 22, 23], this research area is still open.
In this regard, Moore and Geritzer method and the linearized Jacobian matrix as the state-space equations are presented by [24]; but, the role of the positive or negative Real- part of the eigenvalues to determine surge point and compressor instability domains and the issue of the existence of a limit cycle are neglected. This method is developed by considering the square of the rotating stall domain [25]. Although the Jacobian matrix is also utilized to analyse the stability [25], only the Real-part of the eigenvalues is intended while the Imaginary-part is neglected. This paper analyses the aerodynamic equations describing nonlinear state space of compressor by means of Jacobian matrix and phase plane study. Then the results are employed to describe stability and instability domain of compressor characteristic and surge event. In this study, surge embarking criteria and system limit cycle of the surge instability are discussed. To this aim, the compressor map relating inlet to outlet parameters is plotted in different speeds. The compressor characteristic is described as a cubic function of mass flow [26]as follow:

$$\psi_c(\phi) = \psi_{c0} + H\left[1 + 1.5\left(\frac{\phi}{W} - 1\right) - 0.5\left(\frac{\phi}{W} - 1\right)^3\right] \quad (26)$$

Table 1 reveals the designing point of the desired compressor.

**Table 1** Simulation Parameters [26]

| Variable | Value |
|---|---|
| $\psi_{c0}$ | 0.352 |
| H | 0.18 |
| W | 0.25 |

. The compressor characteristic depends on its aerodynamic and its experimental data. "Ψ" and "Φ" are normalized amount of pressure ratio and mass flow rate [27] respectively. "W" and "H" represent the effect of temperature, molecular weight and, compressor geometry. To describe the relation between transient mass flow and pressure ratio, Geritzer equations are given as follows:

$$f1: \quad \dot{\phi} = B \cdot (\psi_c(\phi) - \psi);$$
$$f2: \quad \dot{\psi} = \frac{1}{B} * (\phi - g * \sqrt{\psi}) \quad (27)$$

Where B is a constant parameter, representing rotor rotational speed [28], that is considered equal to 0.8 [29]. "Ψc" displays characteristic curve of the compressor which is obtained in steady states condition. "g" is a constant parameter depends on throttle valve's opening or compressor inlet mass flow rate. In system dynamic equations, g is not related to mass flow and pressure ratio of plenum. In steady states condition, the system's mass flow is equal to inlet mass flow [24]. In all equilibrium points, the following equations are available

$$\psi_c(\phi) = \psi \ ; \ \phi = g*\sqrt{\psi} \ . \tag{28}$$

Simplifying equations in equilibrium condition lead to the following equation $\Psi_c(\phi) = \dfrac{\phi^2}{g^2}$

Equilibrium points of the system are derived by using both the aforementioned equation and an intersection point of characteristic compressor curve ($\Psi_c(\phi)$) with throttle curve ($\dfrac{\phi^2}{g^2}$). Due to the limited working range of compressor (between $0 < \Phi < 0.8$) there is only one equilibrium point for the nonlinear system of a compressor (g is positive constant). Based on part of Extended Jacobian Conjecture for n dimensional ordinary system, It means that using Jacobian matrix to investigate the stability region of system, is a global method. On the other hand, if system is stable with any initial operation point, states of system converge to the equilibrium point. Also if the system is unstable, with any initial operation point, states of system increase and will be infinite. In other words, stability of this system is investigated globally. Moreover, in this paper according to state space model (Greitzer equations), g (varied with opening throttle valve) is the parameter and in fact, it is function of states ("Φ = g√Ψ" and "Ψc (Φ) = Ψ") in the equilibrium point. Then in this conditions, equilibrium point is not constant for various types of g and is dependent on state of the system. Based on the aforementioned statements, eigenvalues of Jacobian matrix are not constant and vary as a function of state (Φ). According to what will be discussed about sign and type of eigenvalue, in present paper investigation of stability covers all range of compressor operation because eigenvalue is a dependent variable varying as a function of mass flow (Φ). Totally, there are one equilibrium point for every constant of g. But by considering g based on Φ, eigenvalues will contain all working region of compressor and there are stable and unstable zones by changing sign and type of eigenvalue.

The partial derivative is used to extract Jacobian matrix as shown in Eq.29,20.

$$\dfrac{\partial f_1}{\partial \phi} = 0.8.\left[0.18*\left(6 - 6.(4\phi - 1)^2\right)\right]; \quad \dfrac{\partial f_1}{\partial \psi} = -0.8 \tag{29}$$

$$\frac{\partial f_2}{\partial \phi} = 1.25; \quad \frac{\partial f_2}{\partial \psi} = 1.25(-g \cdot \frac{1}{2\sqrt{\psi}}) \tag{30}$$

In order to calculate the eigenvalues of Jacobian Matrix, Eq.31 should be solved. In Eq.31 "s" is a symbolic variable and its value represents the eigenvalue of the Jacobian matrix. " I" denotes to the identify matrix. Eq. 29,30 are used to achieve Eq.32 according to Eq.31.

$$det(sI - J) = 0 \tag{31}$$

$$s.s + s\left[0.625 \cdot \frac{g}{\sqrt{\psi}} - 0.864 + 0.864 \cdot (4\phi - 1)^2\right]$$

$$+ 0.625 \cdot \frac{g}{\sqrt{\psi}} \left[-0.864 + 0.864 \cdot (4\phi - 1)^2\right] + 1 = 0 \tag{32}$$

Due to the establishment of equilibrium in Eq.28, these variables "$\Psi_c$" "$\Psi$" and "g", as three unknown variables can be described based on "$\Phi$" as follows:

$$s.s + s.[0.625 \cdot \frac{\phi}{\psi_c(\phi)} - 0.864 + 0.864 \cdot (4\phi - 1)^2]$$

$$+ \frac{0.625 \cdot \phi \cdot [-0.864 + 0.864 \cdot (4\phi - 1)^2]}{\psi_c(\phi)} + 1 = 0 \tag{33}$$

Discriminant of Eq.33 that is described in Eq.34, plays an important role in the dynamic of systems. Eq.34 has negative value in working mass flow range of compressor ($0 < \Phi < 0.8$).

$$\Delta = 0.39 \cdot (\frac{\phi}{\psi_c})^2 + [-0.864 + 0.864 \cdot (4\phi - 1)^2].$$

$$[(-0.864 + .0864 \cdot (4\phi - 1)^2) - 1.25(\frac{\phi}{\psi_c})] - 4 \tag{34}$$

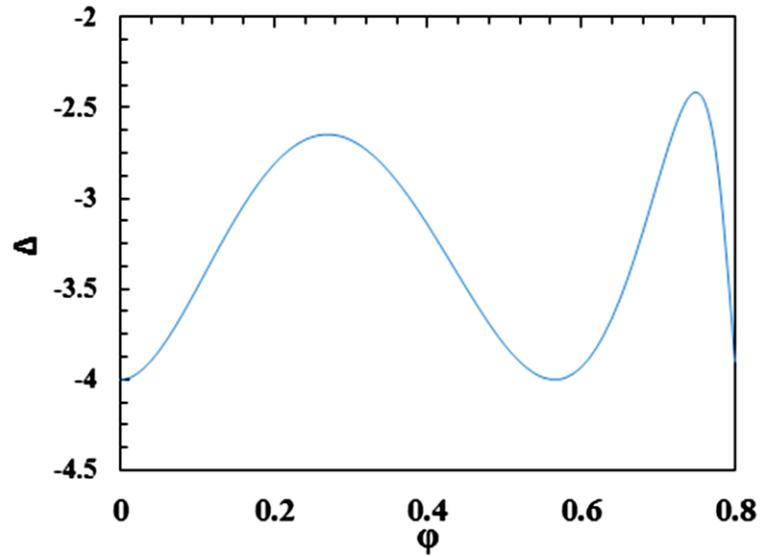

Figure 9 Variations of Δ versus flow rate

Figure 9 represents the variation of discriminant in different mass flow rates. Due to the negative value of Eq.34, the eigenvalues are a complex number. The real part of eigenvalues varies by mass flow according to Eq.35. The sign of the real part of eigenvalues is important to diagnose the stability of operation point.

$$\text{Real part} = -[0.625 \cdot \frac{\phi}{\psi_c(\phi)} - 0.864 + 0.864 \cdot (4\phi - 1)^2]$$

(35)

The real part of eigenvalues varies by mass flow according to Eq.35. The related curve is shown in Fig.10.

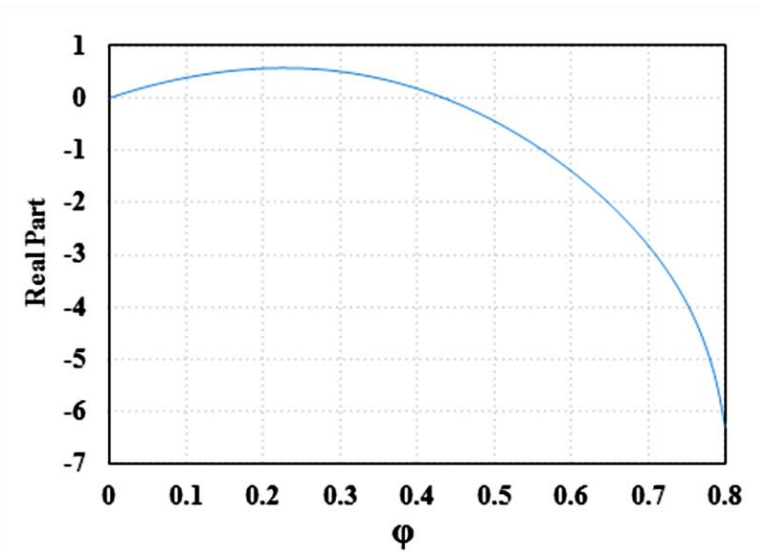

Figure 10 Variations of real part versus flow rate

If the real part of eigenvalues is negative, as an evaluated working point, the working point is a stable focus one. According to what is shown in Fig.10 if the non-dimensional inlet mass flow rate is more than 0.43, the compressor performance is completely stable and far from the surge. In this condition, rotor capability to enforce flow to forward overcomes reverse flow due to positive pressure gradient. The status of compressor's working point on characteristic curve remains stable. If the inlet mass flow varies, according to Greitzer dynamic equations, the working point changes to a new one. Consequently, after transient states the inlet mass flow tends to equilibrium point and the new pressure ratio is obtained. By solving aforementioned equations, the stable zone of intended compressor is determined. Accordingly, for "Φ" amounts more than 0.43, compressor is far from surge. In other words, by considering any initial point with different mass flow rate, the system tends to a new stable statue (new steady states condition on the characteristic curve) by changing working point of compressor and stability is certain. For example, if on the stable statue with "Φ=0.6312" and "Ψ=0.6226", mass flow changes due to a perturbation in compressor system, then both "Φ" and "Ψ" oscillate and tend to the initial statues. Fig.11 represents the system trajectory of transient stability.

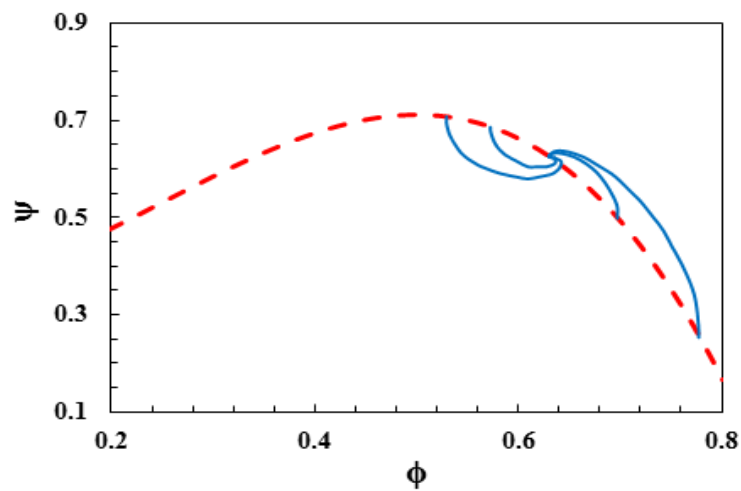

Figure 11 Stable zone of the compressor

Due to the dynamic instability, surge occurs in a mass flow with maximum pressure ratio. Figure 11 illustrates that the maximum value of the pressure ratio is on "$\phi$ =0.5". Since other components of the compressor affects fluid flow inertia, surge appears in lower mass flow rates than the theoretical one. In fact, other components such as pipes and ducts in both inlet and outlet, affect compressor characteristic and its surge margin. On the other hand, flow inertia makes a tendency in the fluid to stay at its condition and resists against variations [30].

## 6.2. Limit cycle or surge phenomenon oscillations

On the other side, since surge causes pressure ratio and mass flow to oscillate, the existence of limit cycle is important to investigate the possibility of an increase in the amplitude of oscillations based on Bendixson theorem [31]. To this purpose, Eq.29,30 are used to extract Eq.36.

$$R = \frac{\partial f_1}{\partial \phi} + \frac{\partial f_2}{\partial \psi} = 0.8.[0.18.(6 - 6.(4\phi - 1)^2)]$$
$$+1.25(-g.\frac{1}{2\sqrt{\psi}}) \tag{36}$$

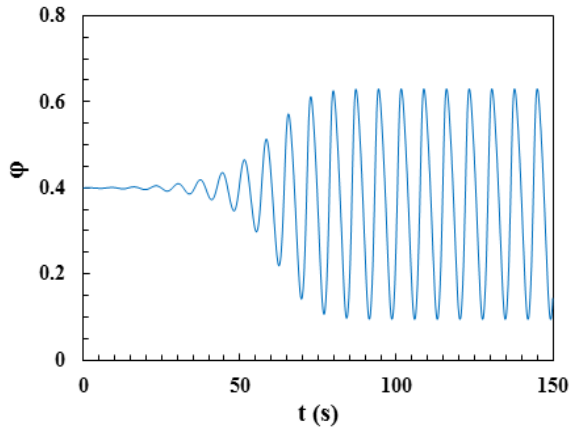

Figure 12 Flow rate variations within unstable region

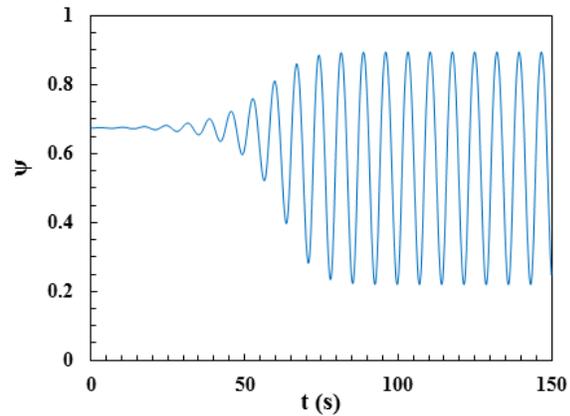

Figure 13 Pressure ratio variations within unstable region

Consequently, according to Bendixson theorem [32], the limit cycle does not exist in stable zone. However, in lower mass flow rates and in an unstable zone, the sign of Eq. 36 changes and limit cycle may exist.
By using the phase diagram, the existence of the limit cycle is proved in an unstable zone. If the working point of compressor or equilibrium point tends to unstable zone, mass flow and pressure ratio variations will oscillate. Despite the fact that the mass flow rate is in an unstable zone, working points which are unstable focus one, tend to limit cycle and amplitude of oscillations remains constant. Therefore, due to the unstable focus inside the limit cycle, the cycle remains stable. In Fig.14, the limit cycle is shown. Analysis of limit cycle (based on this analysis) is also an innovation.

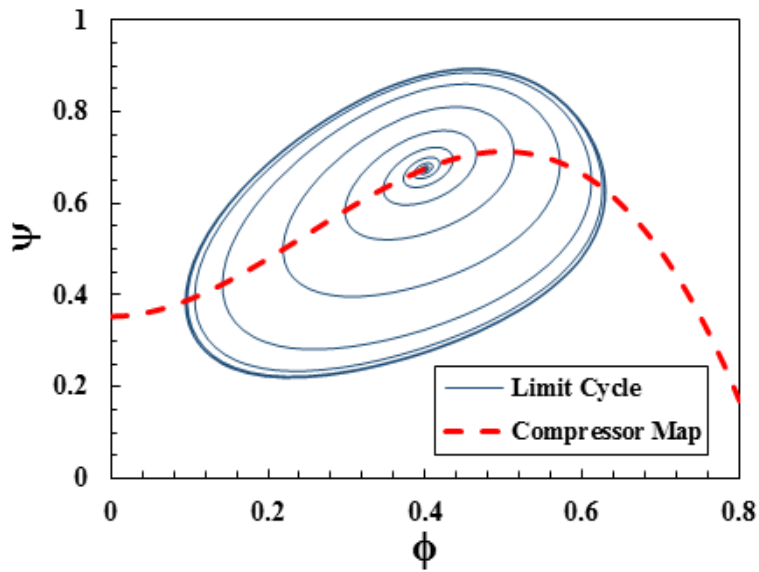

*Figure 14 Limit cycle within unstable region*

## 6.3. Validation of the mathematical method

In order to validate the mathematical method, the mass flow which displays surge inception is compared to what is obtained through experiment by Koff. The case study is a three-stage compressor that for more detail, one can refer to [33]. Koff by using an unstalled method provides a performance map along with anticipated stall point and the axisymmetric peak point. Koff validation method showed that the compressor unstalled performance is for $\Phi > 0.48$. In fact, according to his obtained data, the inception of instabilities is for amounts of "$\Phi$" that are less than 0.48. The related characteristic is shown in Fig.15.

In the present study, despite many simplifying assumptions, the instability inception is about $\Phi < 0.43$. The lower mass flow coefficient which is surge inception is due to many reasons. Firstly, in the utilized mathematical method (Greitzer model), surge and stall inception are considered simultaneously.

Secondly, as shown in Fig.15, there are two concepts of stall point. One concept is a true stall point which is a function of compressor geometry only.

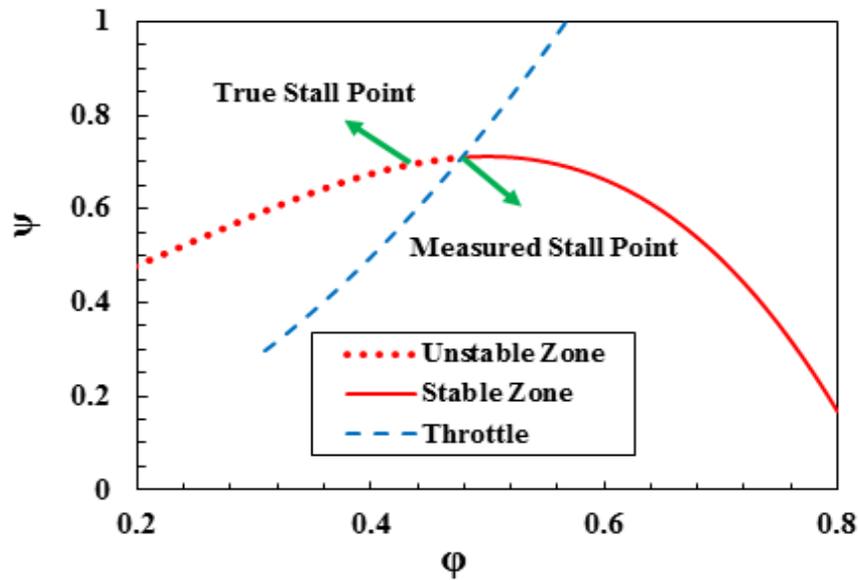

*Figure 15. True and measured stall point*

The other concept of stall point is called measured stall point which exists on a positively sloped portion of characteristic [33]. Since the compression system instability may occur first and then include compressor, these two points are not a coincidence. However, the true stall point occurs in less mass flow coefficient, because it is a function of compressor geometry only. In other words, the true stall point remains hidden from the compressor test. It means that the true stall point is not obtained in the experiment [33]. Moreover, as shown in the characteristic of Fig.15, others suggest that stall can occur at the peak pressure coefficient because the system is dynamically unstable [29]. It was mentioned in [33] that stall is mostly caused by wall stall which is not considered in two-dimensional criteria of maximum pressure coefficient. Consequently, this criterion for the inception of instabilities is a simple moderate prediction.

## 7. Conclusion

This paper has suggested a simple necessary and sufficient condition for the stability and instability analysis of a special class of nonlinear n-dimensional systems, which possess only one equilibrium point. Hartman-Grobman theorem and Popov theorem could prove mentioned claim in this paper. To clarify the issue, 4 examples with different stability statuses have been described and their stability has been analyzed by solving differential equations, plotting the state-space trajectories, their phase portraits and eigenvalues. Results indicated that the rapid consequences from analyzing eigenvalues at the equilibrium point are sufficient, solitary. In addition, the suggested method for the given industrial nonlinear system, illustrated a stability criterion for it.

Analysis of nonlinear compressor system was based on mentioned method, because there was one equilibrium point in this system and validation from experimental data could verify mathematical method. There were two stable and unstable regions for autonomous nonlinear compressor system with changing parameter g(opening throttle valve) and there is only one equilibrium point for every parameter g.